\documentclass[vecphys]{svmult}

\usepackage{makeidx}         
\usepackage{graphicx}        
\usepackage{multicol}        
\usepackage[bottom]{footmisc}

\makeindex             

\begin{document}

\title*{ESO540-032: a Transition-Type Dwarf in the Sculptor Group}
\author{G.\ S.\ Da Costa\inst{1}, H.\ Jerjen\inst{1} \and
A.\ Bouchard\inst{2}}
\institute{Research School of Astronomy \& Astrophysics, ANU, Mt Stromlo 
Observatory, via Cotter Rd, Weston, ACT 2611, Australia.
\texttt{gdc,jerjen@mso.anu.edu.au}
\and Universite de Lyon 1, Centre de Recherche Astronomique de Lyon, 
Observatoire de Lyon, Saint-Genis Laval; and CNRS, École Normale Supérieure 
de Lyon, Lyon, France. \texttt{bouchard@obs.univ-lyon1.fr}}
%
%
\maketitle


\section{Introduction}
\label{sec:1}
The
{\it dwarf Irregular} galaxies (dIrr) are gas-rich systems, with hydrogen mass to blue luminosity ratio 
(M$_{HI}$/L$_{B}$) values exceeding one in solar units, and active star formation.  
The {\it dwarf Elliptical} galaxies (dE), on the other hand, are usually gas-free, with
(M$_{HI}$/L$_{B}$) values substantially less than one.  In most instances they are not
actively forming stars, although their star formation histories are complex
and varied.   These dwarf galaxy types exhibit a morphology-density relation, in the sense that dE 
galaxies tend to be found in denser environments.  For example, within the Local Group, the
vast majority of the dE systems are found as companions to the Milky Way or M31, while the
dIrr members are mostly relatively isolated.  

In the Local Group there are also dwarf galaxies that are {\it transition-types}, systems
that possess moderate amounts of gas (M$_{HI}$/L$_{B}$ $\approx$ 0.1--0.5) and
a low-level of current or recent star formation within a dominant older population.  
We note in passing that the recently discovered distant Milky Way companion 
Leo~T \cite{LeoT} is more likely a low luminosity
dwarf irregular, than a genuine transition-type galaxy, as it appears M$_{HI}$/L$_{B}$ for this dwarf
exceeds one \cite{LeoT}.   The relationship between these classes of dwarf
galaxies remains controversial: e.g., \cite{grebel03} argue that the existence of an offset
in the luminosity-metallicity relation between dEs and dIrrs indicates different evolutionary paths.
They suggest that transition-type dwarfs are the progenitors of dE (dSph)
systems, in the sense that in low density environments where ram-pressure stripping mechanisms
are ineffective, transition-type dwarfs should be common \cite{grebel03}.

One obvious way to test this suggestion is to investigate the properties of dwarf galaxies in
environments beyond the Local Group.  The Sculptor Group is a low density aggregation
of galaxies ranging in distance from $\sim$1.5 to $\sim$4 Mpc. It contains at least five low-luminosity early-type dwarf galaxies.  Their neutral gas content has been studied by 
\cite{bouchard05}, who found that four are likely to be transition-type
objects: the M$_{HI}$/L$_{B}$ values are in the range 0.1--0.2 \cite{bouchard05}.
The fifth system, Scl-dE1, was not detected in HI and has M$_{HI}$/L$_{B}$ $<$ 0.04 
\cite{bouchard05} as expected for a dE system.  The transition-type nature of ESO410-005 and
ESO540-032 is further supported by stellar population studies: both dwarfs 
show evidence for modest amounts of relatively recent star formation \cite{grebel00, jr01}.
We present here new data for ESO540-032: the optical observations reach considerably
fainter than the 
ground-based study of \cite{jr01}, and the new HI observations have higher spatial
resolution and signal-to-noise compared to \cite{bouchard05}.  The transition-type
nature of this dwarf galaxy is confirmed.

\section{Hubble Space Telescope Observations}
\label{sec:2}
HST/ACS observations of ESO540-032 were obtained in August 2006 with total integration times
of 8960sec in the $F606W$ (wide-$V$) filter and 6708sec in the $F814W$ (wide-$I$) filter.
The exposures used a standard dither
pattern and were combined and corrected using the standard ACS
data-processing pipeline. 
The DAOPHOT - ALLSTAR package \cite{Stetson94} was used to determine 
photometry from the combined images.  The
calibration procedures outlined in \cite{Sirianni} were then used to convert the photometry to
the ACS VEGAMAG system and then to Johnson-Cousins $V$ and $I$ magnitudes. 

\begin{figure}[t]
\centering
\includegraphics[height=6.1cm]{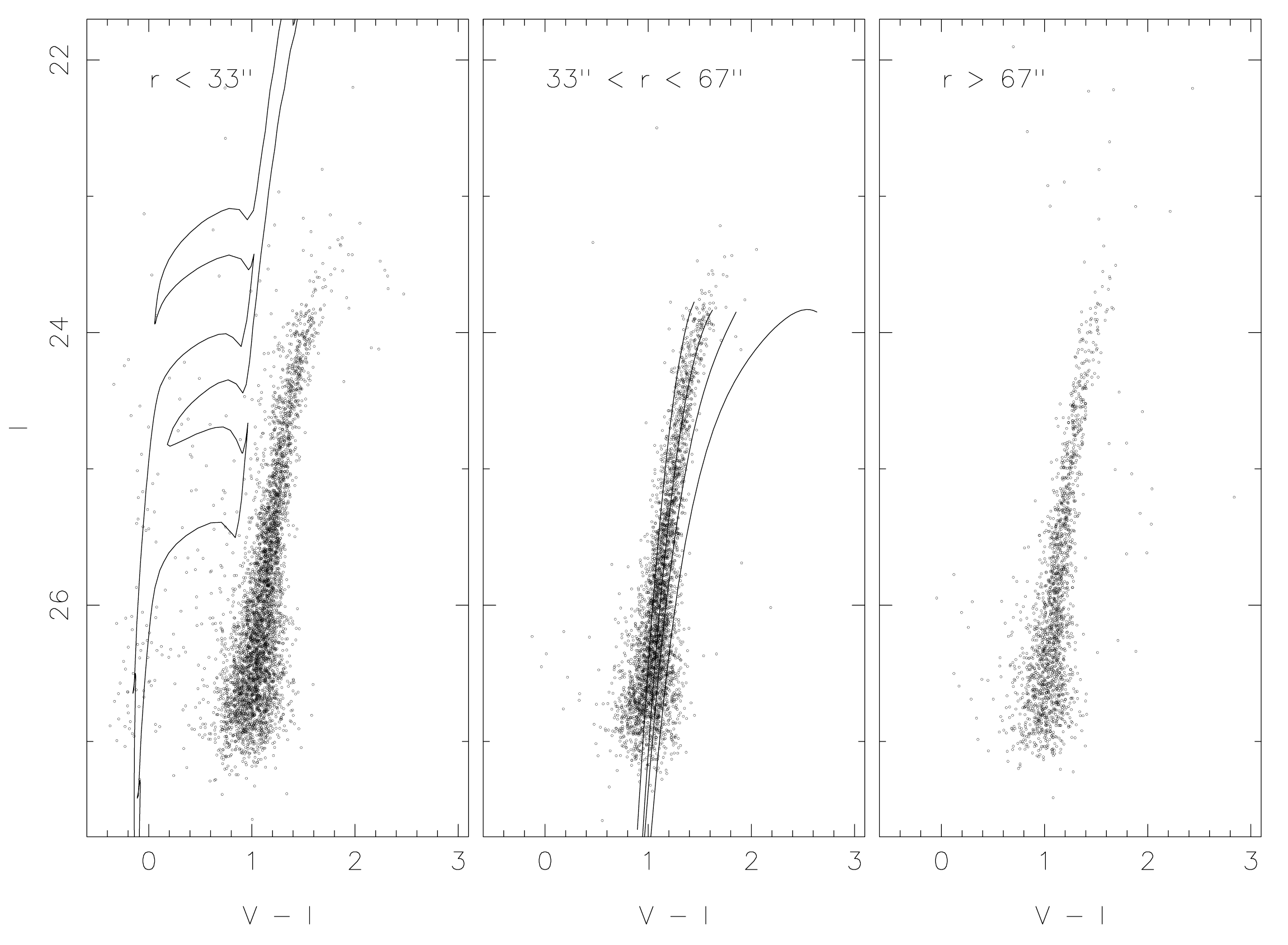}
\caption{Colour-magnitude diagrams for the Sculptor group transition-type dwarf galaxy 
ESO540-032.  Note that the blue stars are found only in the central region.  The 
isochrones shown in the left panel are for ages of 100 and 200Myr and Z=0.001.  The
globular cluster giant branches shown in the middle panel are M15 
([Fe/H]=--2.17), NGC~6752 (--1.54), NGC~1851 (--1.36), and 47~Tuc (-0.71).  Both the isochrones
and the giant branches are plotted using an apparent $I$-band
modulus of 27.87 and a reddening E($V-I$)=0.03. }
\label{fig1}       
\end{figure}

The resulting colour-magnitude diagram is shown in Fig.\  \ref{fig1}, where the data have
been separated into three radial bins based on distance from the galaxy centre.  All three bins
are clearly dominated by an old red giant branch (RGB) population, though as noted originally by
\cite{jr01} there is a small population of blue stars confined to the central regions of 
the galaxy.
Application of a Sobel edge-detection filter to the RGB $I$-band luminosity function constructed 
from Fig.\ \ref{fig1} places the RGB-tip at $I$=23.82$\pm$0.12 mag.  Using a 
reddening of E($V-I$) of 0.03 mag and assuming M$_{I}$(TRGB)=--4.05 (e.g. \cite{DA90}), this 
yields a distance of 3.7$\pm$0.2 Mpc, consistent with previous estimates.   The
luminosity of the dwarf is then M$_{V}$=--12.3, using the total apparent magnitude from \cite{jr01}.
Given the distance estimate, a comparison of the colour of the RGB with those
of globular clusters provides an estimate of the mean metallicity of the dwarf.  This
value, [Fe/H] = --1.7 $\pm$ 0.2, is strictly a lower limit, as at fixed abundance younger RGB stars 
are bluer.  However, since the number of intermediate-age upper-AGB stars above the RGB tip 
in Fig.\ \ref{fig1} is not substantial, the effect of any age difference on the mean metallicity value
is likely to be small.  With this metallicity and the absolute magnitude derived above, the location of ESO540-032 in the metallicity-luminosity diagram is reasonably consistent with the 
Local Group relation.  

Knowledge of the distance also allows isochrones to be fitted to the stars to
the blue of the RGB, the bluest of which undoubtedly represent a main sequence population.
Using isochrones from \cite{Girardi} the age of the blue star population is potentially as
young as 100Myr, perhaps less.  Similar conclusions were reached by 
\cite{jr01}. The isochrones also show that many of the stars
that lie between the main sequence and the RGB,  as well as the stars brighter and bluer than
the RGB tip, can be understood as young stars in post main sequence phases of evolution.
Clearly there has been an episode of star formation in ESO540-032 in the relatively
recent past.  Unfortunately, the small numbers of main sequence stars makes any
attempt to measure the recent star formation history difficult -- deeper observations are
required.

\section{Australia Telescope Compact Array Observations}
\label{sec:3}
\begin{figure}[t]
\centering
\includegraphics[height=5.6cm]{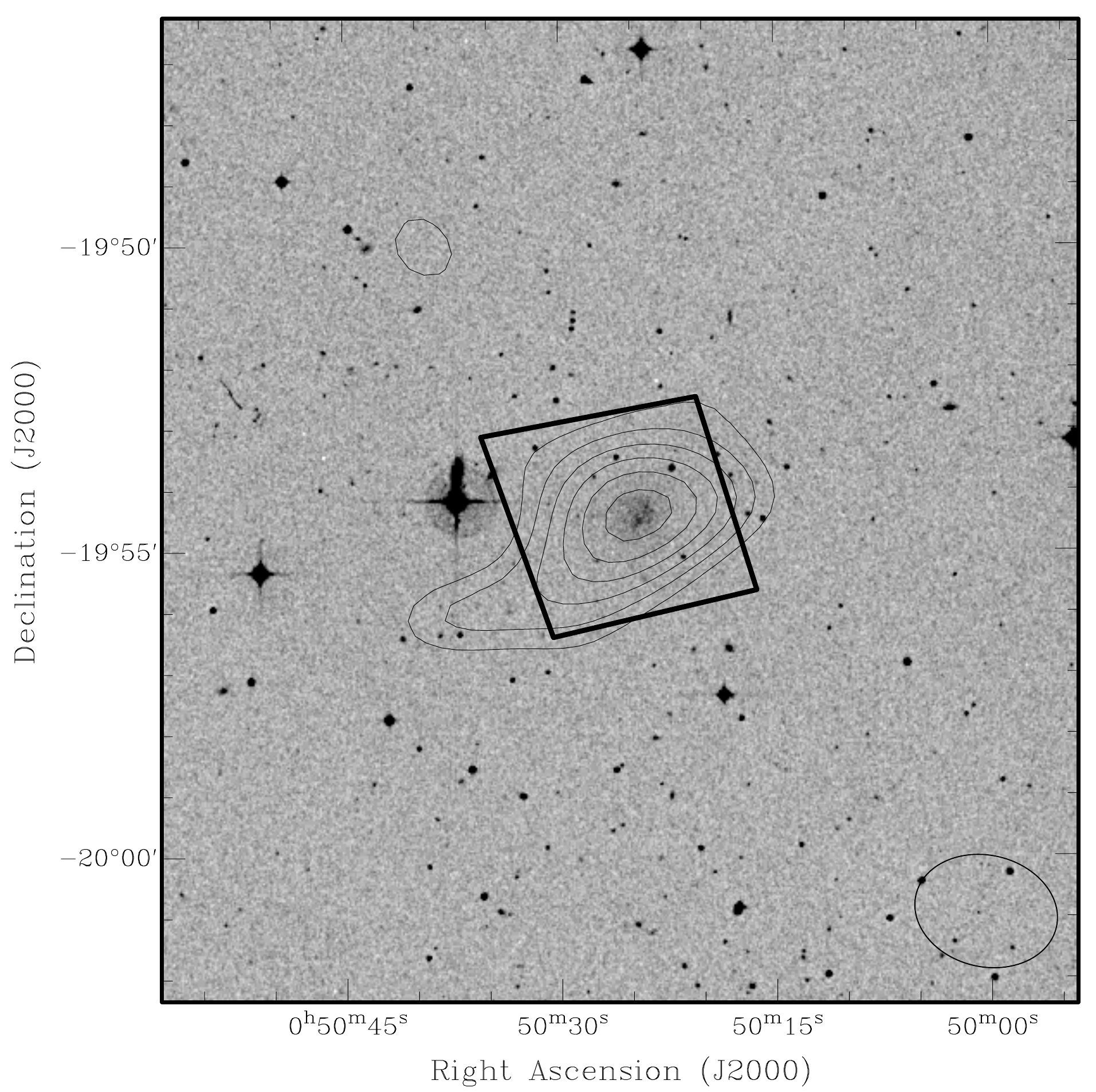}
\caption{HI density contours superposed on a Digital Sky Survey image of ESO540-032.  The lowest 
contour represents $4\times10^{18}$cm$^{-2}$ (2.5$\sigma$ above background) and the contour interval
is $2\times10^{18}$cm$^{-2}$.  The beam size of $109^{\prime\prime}\times142^{\prime\prime}$ is shown
in the bottom right corner.  The quadrilateral outlined by the solid lines is the ACS field.}
\label{fig2}       
\end{figure}

Using the Parkes telescope \cite{bouchard05} detected HI in the field of ESO540-032 at 
a velocity expected
for objects in the Sculptor group.  Using this velocity they were then able to use archival ATCA 
observations to demonstrate the detected HI was spatially coincident with the dwarf galaxy.  The
signal-to-noise of the archival observations, however, was low and the detection marginal.
 To confirm the detection,
and particularly to establish the relative location of the peak of the HI with respect to the optical
image, the galaxy was re-observed using longer integration times.  Fig.\ \ref{fig2} shows the
results: HI is
clearly detected and equally clearly, the HI distribution is centered on the optical
centre of the galaxy.  Further, given the beam size (cf.\ Fig.\ \ref{fig2}), there is 
no compelling evidence to suggest that the HI gas is more extended than the optical light distribution,
which extends at least as far the edge of the ACS field.
The total HI mass is $9.1\pm1.7\times10^{5}$ solar masses, the peak density is 
$15.2\times10^{18}$cm$^{-2}$ and the 
M$_{HI}$/L$_{B}$ ratio is 0.15$\pm$0.04, comparable to Phoenix
 \cite{PhMLR}.

\section{Conclusions}
\label{sec:4}
The presence of a modest amount of recent star formation in an
dominant old stellar population (age at least 6-8 Gyr given the comparative lack of upper-AGB stars),
and the presence of a modest amount of gas, clearly confirm ESO540-032 as a 
transition-type dwarf galaxy in the Sculptor group (cf.\ \cite{jr01,bouchard05}).  We have similar data 
for the other low-luminosity early-type dwarfs in this group -- it will be interesting to see 
whether the analysis of these additional data supports the hypothesis of \cite{grebel03} that
transition-type galaxies will be common compared to dE/dSph galaxies in low density
environments.

%
\index{ESO540-032, dwarf galaxies, Sculptor group}
%



\printindex

\begin{thebibliography}{99.}
\bibitem{bouchard05} A. Bouchard, H. Jerjen, G. S. Da Costa \& J. Ott: AJ, \textbf{130}, 2058 (2005)
\bibitem{DA90} G. S. Da Costa \& T. E. Armandroff: AJ, \textbf{100}, 162 (1990)
\bibitem{Girardi} L. Girardi, G. Bertelli, A. Bressan, C. Choisi et al: A\&A, \textbf{391}, 195 (2002)
\bibitem{grebel03} E. K. Grebel,  J. S. Gallagher \& D. Harbeck: AJ, \textbf{125}, 1926 (2003)
\bibitem{LeoT} M. J. Irwin, V. Belokurov, N. W. Evans  et al:  ApJ, \textbf{656}, L13 (2007)
\bibitem{jr01} H. Jerjen \& M. Rejkuba: A\&A, \textbf{371}, 487 (2001)
\bibitem{grebel00} I. D. Karachenstev, M. E. Sharina, E. K. Grebel et al: ApJ, \textbf{542}, 128 (2000)
\bibitem{Sirianni} M. Sirianni, M. J. Lee, N. Ben\'{i}tez, J. P. Blakeslee et al: PASP, \textbf{117}, 1049 (2005)
\bibitem{Stetson94} P. B. Stetson: PASP, \textbf{106}, 250 (1994)
\bibitem{PhMLR} J. St-Germain, C. Carignan, S. C\^{o}t\'{e}, \& T. Oosterloo: AJ, \textbf{118}, 1235 (1999)
\end{thebibliography}
\end{document}